\documentclass[11pt]{article}
\usepackage{epsfig}
\usepackage{cite}
\def\e{\epsilon}

\newcommand{\be}{\begin{equation}}
\newcommand{\eps}{\epsilon}
\newcommand{\dps}{\displaystyle}
\newcommand{\ee}{\end{equation}}
\newcommand{\bea}{\begin{eqnarray}}
\newcommand{\eea}{\end{eqnarray}}

\newcommand{\loopint}[1]{\int \!\!\! \frac{d^D #1}{\left(2\pi\right)^D}\!}
\newcommand{\ESGamma}{S_{\Gamma}}

\newcommand{\lk}{\left(}
\newcommand{\rk}{\right)}
\newcommand{\lek}{\left[}
\newcommand{\rek}{\right]}
\newcommand{\lp}{\left.}
\newcommand{\rp}{\right.}
\newcommand{\nnb}{\nonumber}

\newcommand{\MB}[2]{\hs{-12} \int\limits_{\hs{15}_{ #1 -i \,
\infty}}^{\hs{15}^{ #1 +i\, \infty}} \hs{-15} \frac{d #2}{2\pi i}}

\newcommand{\hs}[1]{\hspace*{#1 pt}}
\newcommand{\vs}[1]{\vspace*{#1 pt}}

\begin{document}
\begin{titlepage}
\vspace*{-1cm}
\begin{flushright}
PITHA-09/07\\
IPPP/09/09\\
DCPT/09/18\\
TTP09-06\\
SFB/CPP-09-21\\
Saclay-IPhT-T09/021\\
February 2009
\end{flushright}
\vskip 1.9cm

\begin{center}
{\Large\bf Nine-Propagator Master Integrals \\ for Massless Three-Loop Form Factors}
\vskip 1.cm
{\large G.~Heinrich$^{a}$}, {\large T.~Huber$^{b}$}, {\large D.~A.~Kosower$^{c}$}, {\large V.~A.~Smirnov$^{d,e}$}
\vskip .7cm
{\it $^a$Institute for Particle Physics Phenomenology, University of Durham,\\ Durham, DH1 3LE, UK}
\vskip .5cm
{\it $^b$Institut f\"ur Theoretische Physik E, RWTH Aachen University,\\D-52056 Aachen, Germany}
\vskip .5cm
{\it $^c$Institut de Physique Th{\'e}oretique, CEA--Saclay,\\F-91191 Gif-sur-Yvette cedex, France}
\vskip .5cm
{\it $^d$Skobeltsyn Institute of Nuclear Physics of Moscow State University,\\ 119992 Moscow, Russia}
\vskip .5cm
{\it $^e$Institut f{\"u}r Theoretische Teilchenphysik, Universit{\"a}t Karlsruhe (TH),\\D-76128 Karlsruhe, Germany}
\end{center}
\vskip 1cm

\begin{abstract}
We complete the calculation of master integrals for massless
three-loop form factors by computing the previously-unknown three
diagrams with nine propagators in dimensional regularisation. 
Each of the integrals yields a six-fold Mellin-Barnes representation
which we use to compute the coefficients of the Laurent expansion in 
$\epsilon$. Using Riemann $\zeta$ functions of up to weight six, we give
fully analytic results for one integral; for a second, analytic
results for all but the finite term; for the third, analytic results
for all but the last two coefficients in the Laurent expansion.  The
remaining coefficients are given numerically to sufficiently high
accuracy for phenomenological applications.
\end{abstract}
\vfill
\end{titlepage}
\newpage

\section{Introduction}
The quark form factor $\gamma^\ast \to q \bar q$ and gluon form factor
$H \to gg$ (effective coupling) are the simplest processes containing
infrared divergences at higher orders in massless quantum field
theory, and therefore are of particular interest in many aspects.
They have, for instance, been used to predict the infrared pole
structure of multi-leg amplitudes at a given
order~\cite{Magnea:1990zb,catani,sterman,Dixon:2008gr}.  The form
factors can also be exploited to extract resummation
coefficients~\cite{Magnea:2000ss,moch1}, and they enter the purely
virtual corrections to a number of collider reactions (Drell-Yan
process, Higgs production and decay, DIS).

Besides phenomenological applications, a major motivation for
obtaining analytic results at three-loop order and beyond is 
finding and understanding structures in massless gauge theories that
generalize to an arbitrary number of loops.  Much progress has
been achieved recently in the prediction of all-order singularity
structures in QCD~\cite{Becher:2009cu,Gardi:2009qi}, in conjectures
about the all-orders behaviour of maximally supersymmetric Yang-Mills
theories~\cite{Anastasiou:2003kj,Bern:2005iz,Bern:2006vw,Bern:2006ew,%
BES,BESfollowup,Alday:2007hr,WilsonLoops,Alday:2008yw} and in
investigations of the finiteness of N=8
supergravity~\cite{EarlyBernGravity,Green,Bern:2007hh,Bern:2008pv,%
BjerrumBohr:2006yw,Bern:2007xj}.  It
has also been shown, at two
loops~\cite{MertAybat:2006wq,MertAybat:2006mz} and recently even 
for the matter contributions at
three loops~\cite{Dixon:2009gx}, that the soft anomalous dimension
matrix in any massless gauge theory is proportional to the one-loop
matrix.

The two-loop corrections to the massless-quark~\cite{vanneerven} and
gluon~\cite{harlander,ravindran} form factors were computed in
dimensional regularisation with $D=4-2\e$ to order $\e^0$ and
subsequently extended to all orders in $\e$ in ref.~\cite{ghm}.
Two-loop corrections to order $\e^0$ are also available for massive
quarks~\cite{breuther}.  The three-loop form factors to order
$\e^{-1}$ (and $\e^0$ for contributions involving fermion loops in the quark form factor) were
computed in~refs.\cite{moch1,moch2}; see also~ref.\cite{moch4}.

In order to calculate  the quark and gluon form factors
at higher orders in perturbation theory, the amplitudes
are reduced to a small set of master integrals by means of algebraic reduction procedures~\cite{chet,laporta,air,gr,fire}. 
At the three-loop level, the master integrals for massless
form factors were identified in ref.~\cite{cedricpaper}, and results for certain
subsets are available in the
literature~\cite{chet,bekavac,mincer,cedricpaper,Heinrich:2007at}.
Among the three-loop master integrals, the genuine three-loop vertex functions
 are the most challenging ones from a computational point of view.
They correspond to two-particle cuts of the
master integrals for massless four-loop off-shell propagator integrals~\cite{baikov},
which have been used in the calculation of the scalar $R$-ratio~\cite{bck}.
In fact, such a correspondence, via two-particle cuts, between these two
families of master integrals follows from the general result of
ref.~\cite{Baikov:2000jg}.
The derivation of the three-loop vertex
integrals is of comparable complexity to massless four-loop propagator
integrals.

Working in dimensional regularisation and expanding the master
integrals in a Laurent series in $\e$, the finite part of the
three-loop form factors requires the extraction of all coefficients
through (polylogarithmic) weight six\footnote{We prefer to use the
  term ``weight" instead of ``transcendentality", because from a
  mathematical point of view there is no proof that $\zeta_3$ is a
  transcendental number.}, i.e.\ coefficients containing terms up to
$\pi^6$ or $\zeta_3^2$.

Those genuine three-loop vertex functions which contain one-loop or two-loop
propagator insertions were computed in ref.~\cite{cedricpaper}.
The three-loop master integrals which are sufficient in order to obtain the
fermion loop contributions within a Feynman diagrammatic approach were computed in ref.~\cite{Heinrich:2007at}.
The purpose of this Letter is to give the results of the remaining three diagrams which have nine propagators each.
We present analytic results for
all but three coefficients in the $\eps$ expansion, 
along with accurate numerical values for
the remaining ones.

\section{Computational Methods and Results}
\label{sec:results}
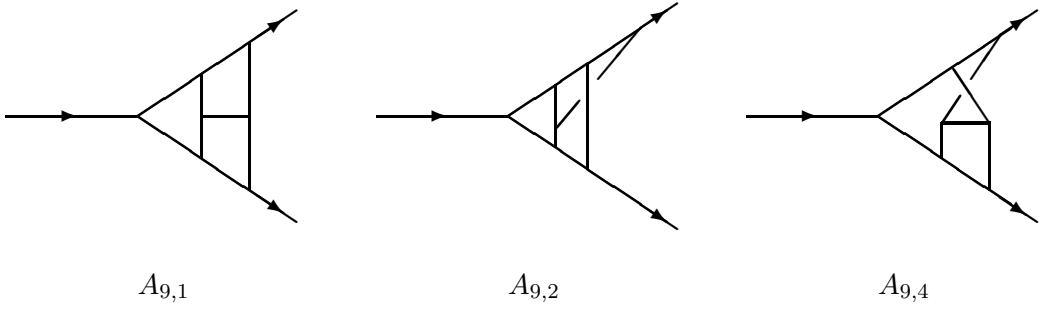
\begin{figure}[t]
    \vs{5}
    \hs{31}
\begin{tabular}{ccc}
\hs{25}
\parbox{4.5cm}{
\begin{picture}(0,0)
\thicklines
\put(-50,0){\vector(1,0){27}}
\put(-23,0){\line(1,0){23}}
\put(0,0){\line(3,2){60}}
\put(0,0){\line(3,-2){60}}
\put(0,0){\vector(3,2){55}}
\put(0,0){\vector(3,-2){55}}
\put(24,-16){\line(0,1){32}}
\put(42,-28){\line(0,1){56}}
\put(24,0){\line(1,0){18}}
\end{picture}\\ \vs{40} \\ $A_{9,1}$}
&
\parbox{4.5cm}{
\begin{picture}(0,0)
\thicklines
\put(-50,0){\vector(1,0){27}}
\put(-23,0){\line(1,0){23}}
\put(0,0){\line(3,2){64}}
\put(0,0){\line(3,-2){64}}
\put(0,0){\vector(3,2){60}}
\put(0,0){\vector(3,-2){60}}
\put(18,-12){\line(0,1){24}}
\put(30,-20){\line(0,1){40}}
\put(18,-5){\line(5,6){9}}
\put(34,14.2){\line(5,6){15.5}}
\end{picture}\\ \vs{40} \\ $A_{9,2}$}
&
 \parbox{4.5cm}{
\begin{picture}(0,0)
\thicklines
\put(-50,0){\vector(1,0){27}}
\put(-23,0){\line(1,0){23}}
\put(0,0){\line(3,2){60}}
\put(0,0){\line(3,-2){60}}
\put(0,0){\vector(3,2){56}}
\put(0,0){\vector(3,-2){56}}
\put(24,-16){\line(0,1){13.4}}
\put(42,-28){\line(0,1){25.6}}
\put(23.8,-2.6){\line(1,0){18.4}}
\put(42,-2.5){\line(-2,3){14}}
\put(24.1,-2.6){\line(2,3){7}}
\put(35.1,13.9){\line(2,3){11.3}}
\end{picture}\\ \vs{40} \\ $A_{9,4}$} \\ \\ \\
  \end{tabular}
    \vs{-20}
  \caption{Three-loop master integrals with nine massless propagators. The incoming momentum is $q=p_1+p_2$. 
  Outgoing momenta are taken to be
  on-shell and massless, $p_1^2=p_2^2=0$.  \label{fig:3loopdiagrams}}
  \end{figure}
In this section, we list the results we obtained for the three-loop
master integrals with nine propagators. They are depicted in
Fig.~\ref{fig:3loopdiagrams} and labeled as in
refs.~\cite{cedricpaper,Heinrich:2007at}.  All other diagrams with up to
eight propagators were already given in the same references.
 We will work in dimensional
regularisation with $D=4-2\e$.
\subsection{Diagram $A_{9,1}$}
The first diagram to be considered is $A_{9,1}$ which can be written as 
follows,
\bea\label{eq:A91}
\dps A_{9,1} &=& \loopint{k}\loopint{l}\loopint{r} \nnb \\
&&\times\frac{1}{\lp k\rp^2 \, \lk k+p_1\rk^2 \, \lk k+l\rk^2  \,
\lk k-r\rk^2 \,\lk l+r\rk^2 \, \lk l+p_2\rk^2\, \lp l \rp^2 \, \lk r+p_1 \rk^2 \, \lk r-p_2\rk^2} \; .
\eea
Here and in the following we tacitly assume that all propagators contain an infinitesimal $+i\eta$. The integral in eq.~(\ref{eq:A91})
can be written in terms of the following six-fold Mellin-Barnes (MB) representation~\cite{smirnov,Tausk,smirnovbook,Alejandro}:
\bea
\dps A_{9,1}&=& i \, \ESGamma^3 \, \lek -\lp q\rp^2-i \, \eta \rek^{-3-3\, \eps} \frac{\Gamma^3(1-\eps)}
{\Gamma(-2\,\eps)} \MB{c_1}{w_1} \MB{c_2}{w_2} \MB{c_3}{w_3} \MB{c_4}{w_4}\MB{c_5}{w_5} \MB{c_6}{w_6}\nnb\\
&&\nnb \\
&& \times \,
\frac{\Gamma(-w_2) \,\Gamma(2+w_1+w_2) \,\Gamma(-w_3) \,\Gamma(w_3-w_2-w_4)\, \Gamma(-w_4) \,\Gamma(-w_5)\,\Gamma(1+w_3+w_5)}
{\Gamma(-w_2-w_4) \,\Gamma(2+w_3+w_5) \,\Gamma(2+w_5+w_6)}\nnb \\
&&\nnb \\
&& \times \, \frac{\Gamma(w_4+w_5-w_1)\,\Gamma(-w_6)\,\Gamma(1+w_5+w_6)\,\Gamma(-2-3\eps-w_3-w_5)}
{\Gamma(-1-4\eps-w_5) \, \Gamma(-1-3\eps-w_2)\, \Gamma(3+\eps+w_1+w_2)\, \Gamma(3+3\eps+w_5)}\nnb \\
&&\nnb \\
&& \times \, \Gamma(-1-2\eps-w_2-w_4)\,\Gamma(-2-3\eps-w_5-w_6) \,\Gamma(2+\eps+w_1+w_2)\, \Gamma(w_1-w_5-\eps) \nnb \\
&& \times   \,\Gamma(3+2\eps+w_2+w_4+w_5+w_6)\,\Gamma(3+3\eps+w_3+w_5)\,\Gamma(3+3\eps+w_5+w_6)\nnb \\
&& \times  \, \Gamma(-1-\eps-w_1)\,\Gamma(-1-\eps-w_2)
\;\; , \label{eq:A91MB}
\eea
\begin{equation}
{\rm where } \qquad q^2=(p_1+p_2)^2 \qquad {\rm and } \qquad S_{\Gamma} = \frac{1}{\lk 4\pi\rk^{D/2}\,\Gamma(1-\e)} \,. \nonumber
\end{equation}
The representation (\ref{eq:A91MB}) was obtained from
eq.~(\ref{eq:A91}) by first introducing Feynman parameters; then,
integrating over the momenta, loop by loop, and finally introducing MB
parameters to decompose sums into products where appropriate.  As
usual, the contour integrals in the complex plane are along curves
which separate left poles of $\Gamma$ functions from right ones, where
``left poles" are poles stemming from a $\Gamma(\ldots +w)$
dependence, while ``right poles" stem from a $\Gamma(\ldots -w)$
dependence~\cite{smirnovbook}.  The most convenient choice for these
contours are straight lines parallel to the imaginary axis, that is
with constant real parts along the curves.  According to
refs.~\cite{Alejandro,Tausk}, these real parts, together with the
parameter $\eps$, must be chosen in such a way as to have positive
arguments in all occurring $\Gamma$ functions in order to separate
left and right poles in the desired way. In certain situations, such
admissible straight contours and an appropriate starting value of $\e$
do not exist. This can be cured via the introduction of an auxiliary
analytic regularisation (see, e.g., refs.~\cite{Alejandro,Tausk}).

The regularisation of the MB integral as described above, as well as
the analytic continuation to $\eps=0$, was done with the {\tt MB}
package~\cite{czakonMB} and, alternatively, with the {\tt MBresolve}
package~\cite{S2MB}, which is based on the strategy formulated in
ref.~\cite{smirnov}.  (Within this latter strategy, straight lines for
the contours along the imaginary axis are not required at the
beginning.)  These packages were also used for numerical cross
checks. Moreover, we have also derived seven-fold MB representations
in two different ways: using the {\tt AMBRE} package~\cite{ambre} and
starting from the MB representation of ref.~\cite{Bern:2005iz} derived
for general powers of the propagators for the tennis court diagram.
The numerical evaluation based on these two MB representations was
again performed with {\tt MB}.  In addition we performed numerical
checks with the sector decomposition methods of~\cite{gudrun1,gudrun2}
and the {\tt FIESTA}~\cite{fiesta} package.  (See ref.~\cite{Bogner:2007cr} for
another implementation of sector decomposition.)

As analytic techniques we apply Barnes's lemmas and the theorem of
residues to the multiple Mellin-Barnes integrals, and insert integral
representations of hypergeometric functions as well as
$\psi$ functions and Euler's $B$ function where appropriate. We also
make use of the {\tt
  HPL}~\cite{Remiddi:1999ew,Gehrmann:2001pz,HPL,HPL2}, {\tt
  HypExp}~\cite{hypexp,Huber:2007dx}, and {\tt
  barnesroutines}~\cite{barnesroutines} packages, as well as an in-house
implementation of the nested sums
algorithm~\cite{Vermaseren:1998uu,nestedsums}. The final result for
$A_{9,1}$ reads,
\bea \dps A_{9,1}&=& i \, \ESGamma^3 \, \lek -\lp
q\rp^2-i \, \eta \rek^{-3-3\, \eps} \nnb \\ &&\times \Big[-\frac{1}{18
    \e^5}+\frac{1}{2 \e^4}+\Big(-\frac{53}{18}-\frac{4 \pi^2}{27}\Big)
  \frac{1}{\e^3}+\Big(\frac{29}{2}+\frac{22
    \pi^2}{27}-2\zeta_3\Big)\frac{1}{\e^2}\nnb\\ && \hs{15}
  +\Big(-\frac{129}{2}-\frac{8\pi^2}{3}+\frac{158}{9}\zeta_3-\frac{20\pi^4}{81}\Big)\frac{1}{\e}\nnb\\ &&
  \hs{15}
  +\Big(\frac{537}{2}+6\pi^2-\frac{578}{9}\zeta_3+\frac{322\pi^4}{405}-\frac{14}{3}\pi^2\zeta_3-\frac{238}{3}\zeta_5\Big)\nnb\\ &&
  \hs{15}
  +\Big(-\frac{2133}{2}-4\pi^2+158\zeta_3-\frac{302\pi^4}{135}-\frac{26}{3}\pi^2\zeta_3+
  \frac{826}{3}\zeta_5-\frac{2398\pi^6}{5103}-\frac{466}{3}\zeta_3^2\Big)\e
  \nnb\\ && \hs{15}+{\cal{O}}(\e^2)\Big] \; .\label{eq:A91expand} 
\label{eq:A91Result}\eea
We emphasize at this point that all terms in eq.~(\ref{eq:A91expand})
have been derived by purely analytic steps.  We also derived this
result by evaluating first the following integral with a numerator:
\bea\label{eq:A91n} \dps A_{9,1}^{(n)} &=&
\loopint{k}\loopint{l}\loopint{r} \nnb \\ &&\times\frac{r^2}{\lp
  k\rp^2 \, \lk k+p_1\rk^2 \, \lk k+l\rk^2 \, \lk k-r\rk^2 \,\lk
  l+r\rk^2 \, \lk l+p_2\rk^2\, \lp l \rp^2 \, \lk r+p_1 \rk^2 \, \lk
  r-p_2\rk^2} \; .  \eea Starting from the above mentioned MB
representation in ref.~\cite{Bern:2005iz} for the tennis court diagram
and setting one of the indices to minus one (the index corresponding
to the numerator $r^2$), we obtained the following seven-dimensional
MB representation: \bea \dps A_{9,1}^{(n)}&=& i \, \ESGamma^3 \, \lek
-\lp q\rp^2-i \, \eta \rek^{-2-3\, \eps} \frac{\Gamma^3(1-\eps)}
{\Gamma(-2\,\eps)} \MB{c_1}{w_1} \MB{c_2}{w_2} \MB{c_3}{w_3}
\MB{c_4}{w_4}\MB{c_5}{w_5} \MB{c_6}{w_6}\MB{c_7}{w_7}\nnb\\ &&\nnb
\\ && \times \,
\frac{ \Gamma(-w_2)\Gamma(1 + w_1 + w_2 + w_3) \Gamma(-w_3) \Gamma(-w_4) \Gamma(1 + w_1 + w_4) \Gamma(-1 - \e - w_1 - w_3)}
{\Gamma(1 - w_2) \Gamma(1 - w_3) \Gamma(1 - 2 \e + w_1 + w_2 + w_3)}\nnb \\
&&\nnb \\
&& \times \, \frac{ \Gamma(-1 - \e - w_1 - w_2 - w_4)\, \Gamma(2 + \e + w_1 + w_2 + w_3 + w_4)\,
  \Gamma(-\e + w_1 + w_3 - w_5)}
{\Gamma(-4 \e - w_5)\, \Gamma(1 - w_6) \,\Gamma(1 - w_4 - w_7) }\nnb \\
&& \nnb \\
&& \times \, \Gamma(2 + 3 \e + w_5)\,
  \Gamma(-1 - 3 \e - w_5 - w_6)\, \Gamma(-w_6)\, \Gamma(1 + w_5 + w_6)  \nnb \\
&& \times   \,
  \Gamma(-1 - 3 \e - w_4 - w_5 - w_7)\, \Gamma(-\e + w_1 + w_2 - w_5 - w_6 - w_7)\,
 \Gamma(-w_5) \,  \Gamma(-w_7)\nnb \\
&& \times  \,  \Gamma(-w_1 + w_5 + w_7) \,\Gamma(1 + \e - w_1 - w_2 - w_3 + w_5 + w_6 + w_7)
\;\; . \label{eq:A91MBwithNum}
\eea
It turns out that at each order in the $\eps$ expansion of
$A_{9,1}^{(n)}$, the coefficients
 have {\sl homogeneous\/} weight.  
This property turns out to be
very helpful when one uses the so-called PSLQ algorithm~\cite{pslq}.
Postulating that a given numerical result can be represented as a linear combination of 
certain constants (typically $\zeta$ values and powers of $\pi$) accompanied by rational coefficients, this algorithm solves for the latter.
Starting from eq.~(\ref{eq:A91MBwithNum}) and
using the {\tt MB} and {\tt MBresolve} packages, we applied the Barnes
lemmas whenever possible to MB integrals which appeared after
expanding in $\eps$.  At this point, at worst two-dimensional MB
integrals were left. We calculated these integrals to an accuracy
of 25 digits, which was sufficient to obtain very stable PSLQ
results.  For the term of highest weight we used the assumption that
it is a linear combination, with rational coefficients, of $\pi^6$
and $\zeta_3^2$. This lead to the following result: 
\bea
\dps A_{9,1}^{(n)}&=& i \, \ESGamma^3 \, \lek -\lp q\rp^2-i \, \eta
\rek^{-2-3\, \eps} \nnb \\ &&\times \Big[-\frac{1}{36
    \e^6}-\frac{\pi^2}{18\e^4}-\frac{14\zeta_3}{9\e^3}-\frac{47\pi^4}{405\e^2}\nnb\\ &&
  \hs{15}
  +\left(-\frac{85}{27}\pi^2\zeta_3-20\zeta_5\right)\frac{1}{\e}
  +\left(-\frac{1160\pi^6}{5103}-\frac{137}{3}\zeta_3^2\right)+{\cal{O}}(\e)\Big]
\; .  
\label{eq:A91nResult}\eea 
Subsequently, we derived an analytic relation between
$A_{9,1}$ and $A_{9,1}^{(n)}$ by means of a Laporta
reduction~\cite{laporta} using, independently, the {\tt
  AIR}~\cite{air} and {\tt FIRE}~\cite{fire} packages. The result
reads\footnote{We also thank Beat T\"odtli for correspondence on this
  point~\cite{Toedtli:2009bn}.}  \bea\label{eq:A91nlaporta} \dps A_{9,1}^{(n)} &=& \frac{8
  (2D-7) (2D-5) (3D-10) (3D-8) (D-3) (1311 D^2-11764
  D+26396)}{9(D-4)^5 (3D-14) (5 D-22) \left(q^2\right)^4} \,
A_{4}\nnb\\ &&+\frac{80 (2D-7) (3D-14) (3D-8) (D-3)^2}{3(D-4)^4
  (5D-22) \left(q^2\right)^3} \, A_{5,1}\nnb\\ &&-\frac{64 (2D-7)
  (D-3)^3 (69 D^2-580 D+1220)}{9(D-4)^4 (3D-14)
  (5D-22)\left(q^2\right)^3} \, A_{5,2} \nnb\\ &&+\frac{8 (3D-14)
  (3D-10) (3D-8) (D-3)^2}{(D-4)^4 (5D-22) \left(q^2\right)^3} \,
A_{5,1}^{(M)}\nnb\\ &&-\frac{32 (2D-7) (D-3)^3 (45 D-202)}{3(D-4)^4
  (5D-22)\left(q^2\right)^3} \, A_{5,2}^{(M)} \nnb\\ &&+ \frac{64
  (2D-7) (D-3)^2}{3(D-4) (3D-14) (5D-22) \left(q^2\right)^2} \,
A_{6,1} - \frac{20 (2D-7) (5D-18)}{9(D-4)^2 \left(q^2\right)^2} \,
A_{6,2}\nnb \\ &&+ \frac{8 (2D-7) (3D-14) (3D-10) (D-3)}{(D-4)^3
  (5D-22) \left(q^2\right)^2} \, A_{6,3}\nnb \\ &&- \frac{2
  (3D-14)}{(5D-22) q^2} \, A_{7,3} - \frac{(3D-14)^2 q^2}{2(D-4)
  (5D-22)} \, A_{9,1} \; .  \eea The integrals $A_{4}$,
$A_{5,1}^{(M)}$, and $A_{5,2}^{(M)}$ are listed in the Appendix, while all
other integrals with fewer than nine propagators are given in
refs.~\cite{Heinrich:2007at,cedricpaper}.  
For convenience, all the corresponding graphs are shown in the Appendix, Fig.~\ref{fig:intslt9}.
We have checked that eqs.~(\ref{eq:A91Result}),(\ref{eq:A91nResult}) satisfy this
relation.

\subsection{Diagram $A_{9,2}$}
The next diagram we consider is $A_{9,2}$. It reads
\bea\label{eq:A92}
\dps A_{9,2} &=& \loopint{k}\loopint{l}\loopint{r} \\
&&\times\frac{1}{\lp k\rp^2 \, \lk k+p_1\rk^2 \, \lk k-l+p_1\rk^2  \,
\lk k-r-l\rk^2 \,\lk l+r\rk^2 \, \lk l+p_2\rk^2\, \lp l \rp^2 \, \lk r+p_1 \rk^2 \, \lk r-p_2\rk^2} \; . \nnb
\eea
Like $A_{9,1}$, it can be written in terms of a six-fold Mellin-Barnes representation.
\bea
\dps A_{9,2}&=& i \, \ESGamma^3 \, \lek -\lp q\rp^2-i \, \eta \rek^{-3-3\, \eps} \frac{\Gamma^3(1-\eps)}
{\Gamma(-2\,\eps)} \MB{c_1}{w_1} \MB{c_2}{w_2} \MB{c_3}{w_3} \MB{c_4}{w_4}\MB{c_5}{w_5} \MB{c_6}{w_6}\nnb\\
&&\nnb \\
&& \times \,
\frac{\Gamma(-w_1) \,\Gamma(2+w_1+w_2) \,\Gamma(-w_5) \,\Gamma(-w_2+w_3+w_4+1)\, \Gamma(w_5-w_2) \,\Gamma(w_5-w_4)}
{\Gamma(1-w_4+w_5) \,\Gamma(-w_1-w_3+w_6) \,\Gamma(1-w_2+w_5+w_6)\, \Gamma(2-\eps+w_1+w_3+w_4)}\nnb \\
&&\nnb \\
&& \times \, \frac{\Gamma(-w_6)\,\Gamma(w_6+1)\,\Gamma(w_6-w_3)\,\Gamma(1+w_5+w_6)\,\Gamma(-2-2\eps-w_1-w_3)}
{\Gamma(2-2\eps+w_1+w_3+w_4) \,\Gamma(-1-w_1-3\eps)\, \Gamma(1-\eps+w_4-w_5)\, \Gamma(3+\eps+w_1+w_2)}\nnb \\
&&\nnb \\
&& \times \, \Gamma(-1-\eps-w_2)\,\Gamma(1-\eps+w_1+w_3) \,\Gamma(w_2-w_4-\eps)\, \Gamma(1+w_4-\eps) \,\Gamma(-1-\eps-w_1)\nnb \\
&& \times   \,\Gamma(1-\eps+w_1+w_3+w_4-w_5-w_6)\,\Gamma(\eps-w_1-w_3-w_4+w_5+w_6)\nnb \\
&& \times  \, \Gamma(w_4-w_5-\eps)\,\Gamma(2+\eps+w_1+w_2)\,\Gamma(3+2\eps+w_1+w_3+w_4)
\;\; . \label{eq:A92MB}
\eea
The techniques we apply are the same as before. The final result for $A_{9,2}$ reads
\bea
\dps A_{9,2}&=& i \, \ESGamma^3 \, \lek -\lp q\rp^2-i \, \eta \rek^{-3-3\, \eps} \nnb \\
&&\times \Big[\frac{2}{9 \e^6}+\frac{5}{6 \e^5}+\Big(-\frac{20}{9}-\frac{7 \pi^2}{27}\Big) \frac{1}{\e^4}+\Big(\frac{50}{9}-\frac{17
\pi^2}{27}-\frac{91}{9}\zeta_3\Big)\frac{1}{\e^3}\nnb\\
&& \hs{15} +\Big(-\frac{110}{9}+\frac{4\pi^2}{3}-\frac{166}{9}\zeta_3-\frac{373\pi^4}{1080}\Big)\frac{1}{\e^2}\nnb\\
&& \hs{15} +\Big(\frac{170}{9}-\frac{16\pi^2}{9}+\frac{494}{9}\zeta_3-\frac{187\pi^4}{540}+\frac{179}{27}\pi^2\zeta_3-167\zeta_5\Big)\frac{1}{\e}\nnb\\
&& \hs{15}
+\left(-670.0785 \pm 0.0326 \right)+{\cal{O}}(\e)\Big] \; .\label{eq:A92expand}
\eea
The number for the finite term was obtained with {\tt
  MB}~\cite{czakonMB}. Again, all pole terms in
eq.~(\ref{eq:A92expand}) have been derived by purely analytic
steps. As in the previous case we performed an independent analytic
calculation of an integral with a numerator, which again turns out to
have homogeneous weight\footnote{We thank Lance Dixon for the
  suggestion that this particular numerator generates homogeneous
  weights.} at each order in the $\eps$ expansion:
\bea\label{eq:A92n}
\dps A_{9,2}^{(n)} &=& \loopint{k}\loopint{l}\loopint{r} \\
&&\times\frac{\lk l-p_1\rk^2}{\lp k\rp^2 \, \lk k+p_1\rk^2 \, \lk k-l+p_1\rk^2  \,
\lk k-r-l\rk^2 \,\lk l+r\rk^2 \, \lk l+p_2\rk^2\, \lp l \rp^2 \, \lk r+p_1 \rk^2 \, \lk r-p_2\rk^2} \; .\nnb
\eea
Again, we relate $A_{9,2}$ to $A_{9,2}^{(n)}$ by means of a Laporta
reduction\cite{air,fire}. The result reads
\bea\label{eq:A92nlaporta}
\dps A_{9,2}^{(n)} &=& -\frac{16 (2D-7) (2D-5) (3D-10) (3D-8) (D-3) (70D-303)}{3(D-4)^4 (2D-9) (5 D-22)
\left(q^2\right)^4} \, A_{4}\nnb\\
&&-\frac{64 (2D-7) (3D-8) (D-3)^2 (19D-84)}{3(D-4)^3 (2D-9) (5D-22) \left(q^2\right)^3} \, A_{5,1}
-\frac{64 (2D-7) (D-3)^3}{3(D-4)^3 (2D-9)\left(q^2\right)^3} \, A_{5,2} \nnb\\
&&-\frac{16 (3D-10) (3D-8) (D-3)^2}{(D-4)^2 (2D-9) (5D-22) \left(q^2\right)^3} \, A_{5,1}^{(M)}
+\frac{64 (2D-7) (3D-10) (D-3)^3}{3(D-4)^3 (2D-9) (5D-22)\left(q^2\right)^3} \, A_{5,2}^{(M)} \nnb\\
&&- \frac{8 (2D-7) (5D-18) (16D-71)}{3(D-4) (2D-9) (5D-22) \left(q^2\right)^2} \, A_{6,2}
+\frac{10 (D-3)}{(2D-9) q^2} \, A_{7,1}\nnb \\
&&-\frac{8 (D-4)^2}{(2D-9) (5D-22) q^2} \, A_{7,3}+ \frac{8 (3D-13) (3D-11)}{(2D-9) (5D-22) q^2} \, \left[ A_{7,4} + A_{7,5}\right]\nnb
\\
&&- \frac{(3D-14)}{(5D-22)} \,\left[ A_{8} - q^2 \, A_{9,2} \right]\; .
\eea
We observed homogeneity of weights for $A_{9,2}^{(n)}$ in all coefficients where we have an analytic result.

To evaluate $A_{9,2}^{(n)}$ we started again with eq.~(\ref{eq:A92MB})
and repeatedly applied the Barnes lemmas as much as possible.  The
resulting integrals were treated numerically. Afterwards, the numbers
were plugged into the right-hand side of eq.~(\ref{eq:A92nlaporta})
and an expansion in $\eps$ was performed.  Assuming homogeneous weight
helped to minimize the number of possible constants in
$A_{9,2}^{(n)}$. There is only one constant each at weight 0,2,3 and 4
(1, $\pi^2$, $\zeta_3$, $\pi^4$) and two constants each at weight 5
($\pi^2\zeta_3$ and $\zeta_5$) and 6 ($\zeta_3^2$ and $\pi^6$).  Using
PSLQ, we reproduced the result in eq.~(\ref{eq:A92expand}) up to order
$1/\eps^2$.  We note that the MB integrals contributing to the finite
part of $A_{9,2}$ have dimensionality as high as five, and therefore
prevent us from achieving an accuracy which is sufficient for a
successful application of the PSLQ algorithm.

For the integral with the numerator, the result reads
\bea
\dps A_{9,2}^{(n)}&=& i \, \ESGamma^3 \, \lek -\lp q\rp^2-i \, \eta \rek^{-2-3\, \eps} \nnb \\
&&\times \Big[-\frac{2}{9 \e^6}+\frac{7\pi^2}{27\e^4}+\frac{91\zeta_3}{9\e^3}+\frac{373\pi^4}{1080\e^2}\nnb\\
&& \hs{15}
+\left(-\frac{179}{27}\pi^2\zeta_3+167\zeta_5\right)\frac{1}{\e}
+\left(395.3405   \pm 0.0326  \right)+{\cal{O}}(\e)\Big] \; .
\eea

\subsection{Diagram $A_{9,4}$}
The last diagram we consider is $A_{9,4}$. It reads
\bea\label{eq:A94}
\dps A_{9,4} &=& \loopint{k}\loopint{l}\loopint{r} \\
&&\times\frac{1}{\lp k\rp^2 \, \lk k+p_1\rk^2 \, \lk k-r\rk^2  \,
\lk k-r-l\rk^2 \,\lk l+r\rk^2 \, \lk l+p_2\rk^2\, \lp l \rp^2 \, \lk r+p_1 \rk^2 \, \lk r-p_2\rk^2} \; . \nnb
\eea
Like the previous integrals, it can be written in terms of a six-fold Mellin-Barnes representation,
\bea
\dps A_{9,4}&=& i \, \ESGamma^3 \, \lek -\lp q\rp^2-i \, \eta \rek^{-3-3\, \eps} \frac{\Gamma^3(1-\eps)}
{\Gamma(-2\,\eps)\Gamma(-1-4\, \eps)} \nnb\\
&& \times\MB{c_1}{w_1} \MB{c_2}{w_2} \MB{c_3}{w_3} \MB{c_4}{w_4}\MB{c_5}{w_5} \MB{c_6}{w_6}\nnb\\
&&\nnb \\
&& \times \,
\frac{\Gamma(-w_1) \,\Gamma(1+w_1+w_2) \,\Gamma(-w_3) \,\Gamma(1-w_1+w_3)\, \Gamma(w_3-w_2) \,\Gamma(1+w_4)\,\Gamma(1+w_5)}
{\Gamma(1-w_1) \,\Gamma(w_1+w_2-w_3-w_4+w_5-2\eps) \,\Gamma(1-2\eps+w_1+w_2)}\nnb \\
&&\nnb \\
&& \times \, \frac{\Gamma(-w_5)\,\Gamma(w_4-w_5+1)\,\Gamma(w_5-w_4)\,\Gamma(-w_6)\,\Gamma(1+w_3+w_4+w_6-w_5)}
{\Gamma(2-w_1+w_3+w_4) \,\Gamma(1-w_2+w_3+w_4-w_5)\, \Gamma(2+w_3+w_4+w_6)}\nnb \\
&&\nnb \\
&& \times \, \Gamma(-2-3\eps-w_4)\,\Gamma(w_1+w_2-w_3-2\eps) \,\Gamma(-w_1-\eps)\, \Gamma(w_1-\eps) \,\Gamma(-1-\eps-w_2)\nnb \\
&& \times   \,\Gamma(-2-3\eps-w_3-w_4+w_5-w_6)\,\Gamma(1+\eps-w_1-w_2+w_3+w_4)\nnb \\
&& \times  \, \Gamma(1+w_2-\eps)\,\Gamma(2+\eps+w_1+w_2+w_6)\,\Gamma(3+3\eps+w_3+w_4+w_6)
\;\; . \label{eq:A94MB}
\eea
The techniques we apply are the same as above. 
The final result for $A_{9,4}$ reads,
\bea
\dps A_{9,4}&=& i \, \ESGamma^3 \, \lek -\lp q\rp^2-i \, \eta \rek^{-3-3\, \eps} \nnb \\
&&\times \Big[\frac{1}{9 \e^6}+\frac{8}{9 \e^5}+\Big(-1-\frac{10 \pi^2}{27}\Big) \frac{1}{\e^4}+\Big(-\frac{14}{9}-\frac{47
\pi^2}{27}-12\zeta_3\Big)\frac{1}{\e^3}\nnb\\
&& \hs{15} +\Big(17+\frac{71\pi^2}{27}-\frac{200}{3}\zeta_3-\frac{47\pi^4}{810}\Big)\frac{1}{\e^2}\nnb\\
&& \hs{15} +\Big(117.3999538\pm 0.0000032\Big)\frac{1}{\e}\nnb\\
&& \hs{15}
+\left(1948.167043 \pm 0.000025 \right)+{\cal{O}}(\e)\Big] \; .\label{eq:A94expand}
\eea
The two numbers were again obtained with {\tt MB}~\cite{czakonMB}.
All higher pole terms in eq.~(\ref{eq:A94expand}) have again been
derived by purely analytic steps. In the case of $A_{9,4}$ 
a homogeneous-weight
master integral also exists, with a numerator:
\bea\label{eq:A94n}
\dps A_{9,4}^{(n)} &=& \loopint{k}\loopint{l}\loopint{r} \\
&&\times\frac{r^2}{\lp k\rp^2 \, \lk k+p_1\rk^2 \, \lk k-r\rk^2  \,
\lk k-r-l\rk^2 \,\lk l+r\rk^2 \, \lk l+p_2\rk^2\, \lp l \rp^2 \, \lk r+p_1 \rk^2 \, \lk r-p_2\rk^2} \; .\nnb
\eea
For $A_{9,4}$, we also used the same alternative method of calculation as in the
case of $A_{9,2}$. This procedure was again based on a relation
between $A_{9,4}$ to $A_{9,4}^{(n)}$, which follows from a Laporta
reduction\cite{air,fire}. The relation reads
\bea\label{eq:A94nlaporta}
\dps A_{9,4}^{(n)} &=& -\frac{16 (2D-7) (2D-5) (3D-10) (3D-8) (D-3) (348D^2-3037D+6618)}{9(D-4)^4 (2D-9) (3D-14) (5 D-22)
\left(q^2\right)^4} \, A_{4}\nnb\\
&&+\frac{32 (2D-7) (3D-8) (D-3)^2 (31D-138)}{3(D-4)^3 (2D-9) (5D-22) \left(q^2\right)^3} \, A_{5,1} \nnb\\
&&+\frac{128 (2D-7) (D-3)^3 (195 D^2-1726 D+3816)}{9(D-4)^3 (2D-9) (3D-14) (5 D-22)\left(q^2\right)^3} \, A_{5,2} \nnb\\
&&- \frac{128 (2D-7) (D-3)^2}{3(D-4) (3D-14) (5D-22) \left(q^2\right)^2} \, A_{6,1}
+ \frac{8 (2D-7) (5D-18) (28D-123)}{9(D-4) (2D-9) (5D-22) \left(q^2\right)^2} \, A_{6,2}
\nnb \\
&&-\frac{8 (2D-7) (3D-10) (D-3) (7D-30)}{(D-4)^2(2D-9) (5D-22) \left(q^2\right)^2} \, A_{6,3}+ \frac{20 (D-3)}{(2D-9)q^2} \, A_{7,2} \nnb
\\
&&-\frac{8 (3D-13)(3D-11)}{(2D-9)(5D-22)q^2} \, \left[ A_{7,4} + A_{7,5}\right]- \frac{(3D-14)^2 q^2}{2(2D-9)(5D-22)} \, A_{9,4} \; .
\eea
For the integral with numerator, the
result reads
\bea
\dps A_{9,4}^{(n)}&=& i \, \ESGamma^3 \, \lek -\lp q\rp^2-i \, \eta \rek^{-2-3\, \eps} \nnb \\
&&\times \Big[\frac{1}{9 \e^6}-\frac{10\pi^2}{27\e^4}-\frac{12\zeta_3}{\e^3}-\frac{47\pi^4}{810\e^2}\nnb\\
&& \hs{15}
+\left(206.7612077   \pm 0.0000032  \right)\frac{1}{\e}
+\left( 1237.300592 \pm 0.000035  \right)+{\cal{O}}(\e)\Big] \; .
\eea
As was the case for $A_{9,2}$, the remaining Mellin-Barnes integrals
are of too high dimension to allow for a stable PSLQ fit to determine 
the remaining coefficients analytically.
However, we are confident that a dedicated effort to determine the remaining 
coefficients analytically will eventually be successful.

\section{Conclusions and Outlook}\label{sec:conc}

In this Letter we have evaluated the three nine-propagator master
integrals needed for computing the quark and gluon form factors to
three-loop order.
Each of the three integrals can be expressed in terms of a six-fold
Mellin-Barnes representation from which we determine all coefficients
through weight six in the Riemann $\zeta$ function.  One integral is
given fully analytically up to order $\eps$.  For the second one we
give an analytic representation for all pole parts, and the third one
is given analytically except for the coefficients of $1/\eps$ and
$\eps^0$ in the Laurent expansion in $\eps$.  The remaining Laurent
coefficients are given numerically to an accuracy which is sufficient
for all phenomenological applications. 

\vspace{4mm}

\noindent{\bf Note added:} Our results for the coefficients of the three master integrals $A_{9,1}$,
$A_{9,2}$ and $A_{9,4}$ partially overlap with those of ref.~\cite{BCSSS}
where these integrals were evaluated in an indirect way.
Agreement has been found for all common coefficients:
terms up to $\eps^0$ for $A_{9,1}$, up to $1/\eps$ for $A_{9,2}$
and up to $1/\eps^2$ for $A_{9,4}$. To avoid confusion, we would like to point out 
that the convention for the overall prefactor used in ref.~\cite{BCSSS} differs from our one.
For other recent progress on the three-loop quark form factor see ref.~\cite{Toedtli:2009bn}.

\section*{Acknowledgements}

We would like to thank Thomas Gehrmann for useful correspondence.
This work was supported by Deutsche Forschungsgemeinschaft, SFB/TR 9
``Computergest\"{u}tzte Theoretische Teilchenphysik'', by the German
Federal Ministry of Education and Research (BMBF), by the UK Science
and Technology Facilities Council, and by the Russian Foundation for
Basic Research, grant 08-02-01451. DAK and VAS also acknowledge the support of the ECO-NET program of the
{\sc Egide} under grant 12516NC as
well as the hospitality of the Galileo Galilei Institute in Florence,
during its workshop, ``Advancing Collider Physics: from Twistors to Monte
Carlos'' (August -- October 2007) where part of this work was carried out.
TH acknowledges hospitality from the CERN theory group, where part
of this work was performed.

\appendix

\section{Additional integrals}

\begin{figure}[t]
    \vs{5}
    \hs{31}
\begin{tabular}{ccc}
\hs{25}
\parbox{4.cm}{
\begin{picture}(0,0)
\thicklines
\put(-50,0){\vector(1,0){20}}
\put(-30,0){\line(1,0){18}}
\put(8,0){\circle{40}}
\put(28,0){\vector(1,0){20}}
\put(48,0){\line(1,0){18}}
\qbezier(-12,0)(8,-20)(28,0)
\qbezier(-12,0)(8,20)(28,0)
\end{picture}\\ \vs{25} \\ $A_{4}$}
&
\parbox{4.cm}{
\begin{picture}(0,0)
\thicklines
\put(-50,0){\vector(1,0){19}}
\put(-38,0){\line(1,0){17}}
\put(-5,0){\circle{30}}
\put(27,0){\circle{30}}
\put(11,0){\vector(1,0){46}}
\put(48,0){\line(1,0){18}}
\end{picture}\\ \vs{25} \\ $A_{5,1}^{(M)}$}
&
 \parbox{4.cm}{
\begin{picture}(0,0)
\thicklines
\put(-50,0){\vector(1,0){19}}
\put(-38,0){\line(1,0){17}}
\put(-25,0){\line(1,0){70}}
\qbezier(-21,0)(-21,15)(-6,15)
\qbezier(-6,15)(9,15)(9,0)
\qbezier(9,0)(9,15)(24,15)
\qbezier(24,15)(39,15)(39,0)
\qbezier(-21,0)(-21,-30)(9,-30)
\qbezier(9,-30)(39,-30)(39,0)
\put(39,0){\vector(1,0){15}}
\put(48,0){\line(1,0){18}}
\end{picture}\\ \vs{25} \\ $A_{5,2}^{(M)}$} \\ \\ \\ \\
  \end{tabular}
    \vs{20}
    \hs{31}
  \begin{tabular}{cccc}
  \parbox{3.3cm}{
\begin{picture}(0,0)
\thicklines
\put(-30,0){\vector(1,0){17}}
\put(-13,0){\line(1,0){13}}
\put(0,0){\line(3,2){40}}
\put(0,0){\line(3,-2){40}}
\put(0,0){\vector(3,2){37}}
\put(0,0){\vector(3,-2){37}}
\put(24,-16){\line(0,1){32}}
\qbezier[80](24,-15.9)(34,0)(24,16)
\qbezier[80](24,-16.1)(12.5,0)(24,16)
\end{picture}\\ \vs{30} \\ $A_{5,1}$}
  & 
\parbox{3.3cm}{
\begin{picture}(0,0)
\thicklines
\put(-30,0){\vector(1,0){17}}
\put(-13,0){\line(1,0){13}}
\put(0,0){\line(3,2){40}}
\put(0,0){\line(3,-2){40}}
\put(0,0){\vector(3,2){37}}
\put(0,0){\vector(3,-2){37}}
\put(24,-16){\line(0,1){32}}
\qbezier[80](24,-16)(34,0)(24,16)
\qbezier[80](0,0)(8,-20)(24,-16)
\end{picture}\\ \vs{30} \\ $A_{5,2}$}
  & 
\parbox{3.3cm}{
\begin{picture}(0,0)
\thicklines
\put(-30,0){\vector(1,0){17}}
\put(-13,0){\line(1,0){13}}
\put(0,0){\line(3,2){40}}
\put(0,0){\line(3,-2){40}}
\put(0,0){\vector(3,2){37}}
\put(0,0){\vector(3,-2){37}}
\qbezier[60](24,16.3)(18,12)(24,0)
\qbezier[60](24,-16.3)(18,-12)(24,0)
\qbezier[60](24,16.3)(30,12)(24,0)
\qbezier[60](24,-16.3)(30,-12)(24,0)
\end{picture}\\ \vs{30} \\ $A_{6,1}$}
  &
\parbox{3.3cm}{
\begin{picture}(0,0)
\thicklines
\put(-30,0){\vector(1,0){17}}
\put(-13,0){\line(1,0){13}}
\put(0,0){\line(3,2){40}}
\put(0,0){\line(3,-2){40}}
\put(0,0){\vector(3,2){37}}
\put(0,0){\vector(3,-2){37}}
\put(27,-18){\line(0,1){36}}
\put(26.8,-17.8){\line(-1,3){5.9}}
\put(26.8,17.8){\line(-1,-3){5.9}}
\put(0,0){\line(1,0){20.8}}
\end{picture}\\ \vs{30} \\ $A_{6,2}$} \\ \\ \\ \\ \\
  \end{tabular}
  \hs{50}
  \begin{tabular}{ccc}
  \parbox{4.5cm}{
\begin{picture}(0,0)
\thicklines
\put(-50,0){\vector(1,0){27}}
\put(-23,0){\line(1,0){23}}
\put(0,0){\line(3,2){60}}
\put(0,0){\line(3,-2){60}}
\put(0,0){\vector(3,2){57}}
\put(0,0){\vector(3,-2){57}}
\put(46.2,-30.8){\line(0,1){61.6}}
\put(18,12){\line(2,-3){28.7}}
\qbezier[80](17.6,12.9)(20,40)(46,31)
\end{picture}\\ \vs{40} \\ $A_{6,3}$}
 &
\parbox{4.5cm}{
\begin{picture}(0,0)
\thicklines
\put(-50,0){\vector(1,0){27}}
\put(-23,0){\line(1,0){23}}
\put(0,0){\line(3,2){60}}
\put(0,0){\line(3,-2){60}}
\put(0,0){\vector(3,2){55}}
\put(0,0){\vector(3,-2){55}}
\put(24,-16){\line(2,5){17.4}}
\put(42,-28){\line(-1,2){10.5}}
\put(27,2){\line(-1,2){6}}
\qbezier[60](0,0)(0,26)(20.7,14)
\end{picture}\\ \vs{40} \\ $A_{7,1}$}
  &   
  \parbox{4.5cm}{
\begin{picture}(0,0)
\thicklines
\put(-50,0){\vector(1,0){27}}
\put(-23,0){\line(1,0){23}}
\put(0,0){\line(3,2){60}}
\put(0,0){\line(3,-2){60}}
\put(0,0){\vector(3,2){55}}
\put(0,0){\vector(3,-2){55}}
\put(24,-16){\line(2,5){17.4}}
\put(42,-28){\line(-1,2){10.5}}
\put(27,2){\line(-1,2){6}}
\qbezier[60](20.7,14)(20,40)(41,27.7)
\end{picture}\\ \vs{40} \\ $A_{7,2}$}    \\ \\ \\ \\ \\
\parbox{4.5cm}{
\begin{picture}(0,0)
\thicklines
\put(-50,0){\vector(1,0){27}}
\put(-23,0){\line(1,0){23}}
\put(0,0){\line(3,2){60}}
\put(0,0){\line(3,-2){60}}
\put(0,0){\vector(3,2){57}}
\put(0,0){\vector(3,-2){57}}
\put(46.2,-30.8){\line(0,1){61.6}}
\put(18,12){\line(2,-3){28.7}}
\put(18,-12){\line(0,1){24}}
\end{picture}\\ \vs{40} \\ $A_{7,3}$}
 &
 \parbox{4.5cm}{
\begin{picture}(0,0)
\thicklines
\put(-50,0){\vector(1,0){27}}
\put(-23,0){\line(1,0){23}}
\put(0,0){\line(3,2){60}}
\put(0,0){\line(3,-2){60}}
\put(0,0){\vector(3,2){57}}
\put(0,0){\vector(3,-2){57}}
\put(18,-12){\line(0,1){24}}
\put(18,-12){\line(2,3){6.7}}
\put(28,3){\line(2,3){18.5}}
\put(18,12){\line(2,-3){28.7}}
\end{picture}\\ \vs{40} \\ $A_{7,4}$}
&
\parbox{4.5cm}{
\begin{picture}(0,0)
\thicklines
\put(-50,0){\vector(1,0){27}}
\put(-23,0){\line(1,0){23}}
\put(0,0){\line(3,2){60}}
\put(0,0){\line(3,-2){60}}
\put(0,0){\vector(3,2){57}}
\put(0,0){\vector(3,-2){57}}
\put(46.2,-30.8){\line(0,1){61.6}}
\put(18,-12){\line(2,3){6.7}}
\put(28,3){\line(2,3){18.5}}
\put(18,12){\line(2,-3){28.7}}
\end{picture}\\ \vs{40} \\ $A_{7,5}$} \\ \\ \\ \\ \\
 \parbox{4.5cm}{
\hs{45}
\begin{picture}(0,0)
\thicklines
\put(-50,0){\vector(1,0){27}}
\put(-23,0){\line(1,0){23}}
\put(0,0){\line(3,2){64}}
\put(0,0){\line(3,-2){64}}
\put(0,0){\vector(3,2){60}}
\put(0,0){\vector(3,-2){60}}
\put(30,-20){\line(0,1){40}}
\put(30,-20){\line(-4,5){20.7}}
\put(18,-5){\line(5,6){9}}
\put(34,14.2){\line(5,6){15.5}}
\end{picture}\\ \vs{35} \\\hs{35} $A_{8}$}
  &   
\parbox{4.5cm}{
\hs{60}
\begin{picture}(0,0)
\thicklines
\put(-50,0){\vector(1,0){27}}
\put(-23,0){\line(1,0){23}}
\put(0,0){\line(3,2){60}}
\put(0,0){\line(3,-2){60}}
\put(0,0){\vector(3,2){56}}
\put(0,0){\vector(3,-2){56}}
\put(18,-12){\line(2,3){28.5}}
\put(18,12){\line(2,-3){28.5}}
\end{picture}\\ \vs{35} \\\hs{55} $A_{8,B}$} \\ \\ \\
  \end{tabular}
  \caption{Graphs with fewer than nine propagators. The incoming momentum is $q=p_1+p_2\, , \, p_1^2=p_2^2=0$. \label{fig:intslt9}}
  \end{figure}
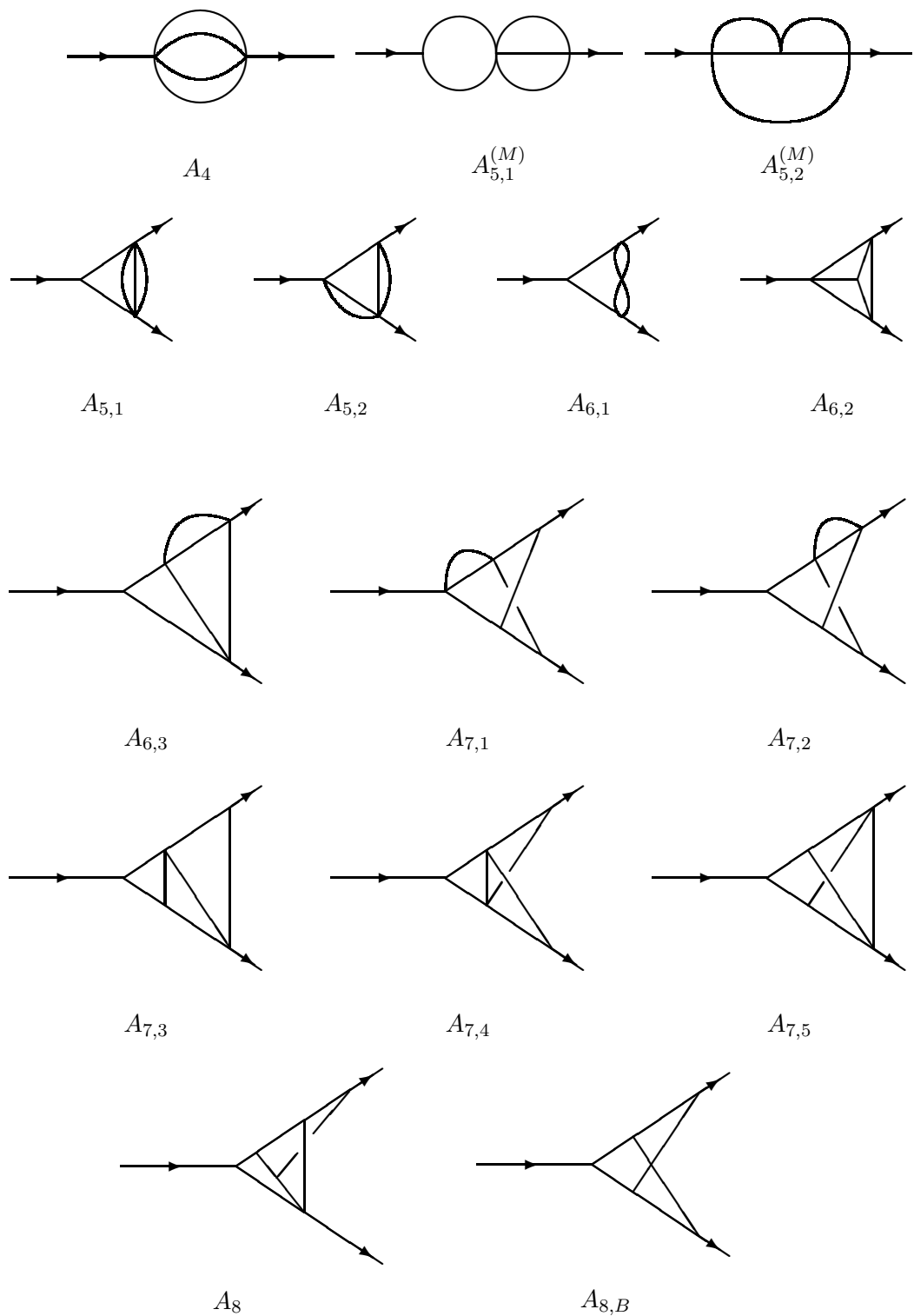
  
In this appendix we collect some additional integrals which appear in the Laporta reduction of the nine-propagator integrals with numerator.
Three of the integrals are two-point functions of the MINCER type~\cite{mincer}. 
The fourth one is the eight-propagator butterfly graph. 
The diagrams are depicted in Fig.~\ref{fig:intslt9}, together with 
diagrams which have been calculated in Ref.~\cite{Heinrich:2007at}.
The results for the two-point functions are
\bea\label{eq:A4}
\dps A_{4} &=& \loopint{k}\loopint{l}\loopint{r}\; \frac{1}{ \lk k+q\rk^2 \, \lk k+l\rk^2  \, \lk l+r\rk^2 \, \lp r \rp^2 } \nnb\\
           &=& -i \, \ESGamma^3 \, \lek -\lp q\rp^2-i \, \eta \rek^{2-3\, \eps} \, \frac{\Gamma^7(1-\eps)
       \Gamma(-2+3\eps)}{\Gamma(4-4\eps)} \; ,
\eea
\bea\label{eq:AM51}
\dps A_{5,1}^{(M)} &=& \loopint{k}\loopint{l}\loopint{r} \;\frac{1}{ \lp k \rp^2\lk k+q\rk^2 \, \lk l+q\rk^2  \, \lk l+r\rk^2 \, \lp r \rp^2 } \nnb\\
           &=& i \, \ESGamma^3 \, \lek -\lp q\rp^2-i \, \eta \rek^{1-3\, \eps} \, \frac{\Gamma^8(1-\eps) \Gamma(\eps)
       \Gamma(-1+2\eps)}{\Gamma(2-2\eps)\Gamma(3-3\eps)}  \; ,
\eea
\bea\label{eq:AM52}
\dps A_{5,2}^{(M)} &=& \loopint{k}\loopint{l}\loopint{r} \; \frac{1}{ \lk r+q\rk^2 \, \lp k \rp^2 \, \lp l \rp^2 \lk k+r\rk^2  \, \lk l+r\rk^2 } \nnb\\
           &=& i \, \ESGamma^3 \, \lek -\lp q\rp^2-i \, \eta \rek^{1-3\, \eps} \, \frac{\Gamma^8(1-\eps) \Gamma^2(\eps)
       \Gamma(-1+3\eps)\Gamma(2-3\eps)}{\Gamma^2(2-2\eps)\Gamma(2\eps)\Gamma(3-4\eps)}  \; .
\eea

\noindent The last integral which we give is the eight-propagator butterfly graph $A_{8,B}$, 
depicted in Fig.~\ref{fig:intslt9}. It is not a master integral, but
it is useful in quite a number of calculations. It is obtained from $A_{9,1}$ by shrinking the horizontal propagator to a point. From
the Laporta reduction we obtain
\bea\label{eq:A8B}
\dps A_{8,B} &=& \loopint{k}\loopint{l}\loopint{r} \nnb \\
&&\times\frac{1}{\lp k\rp^2 \, \lk k+p_1\rk^2 \, 
\lk k-r\rk^2 \,\lk l+r\rk^2 \, \lk l+p_2\rk^2\, \lp l \rp^2 \, \lk r+p_1 \rk^2 \, \lk r-p_2\rk^2} \nnb \\
&&\nnb\\
&=& -\frac{128 (2D-7) (D-3)^3}{(D-4)^3 (3D-14) \left(q^2\right)^3} \, A_{5,2}
+ \frac{32 (2D-7) (D-3)^2}{3(D-4)^2 (3D-14) \left(q^2\right)^2} \, A_{6,1} \nnb \\
&&\nnb\\
&&+ \frac{48 (2D-7) (2D-5) (3D-10) (3D-8) (D-3)}{(D-4)^4 (3D-14) \left(q^2\right)^4} \, A_{4} \; .
\eea
Hence the integral can be written entirely in terms of $\Gamma$ functions. Its expansion reads
\bea
\dps A_{8,B}&=& i \, \ESGamma^3 \, \lek -\lp q\rp^2-i \, \eta \rek^{-2-3\, \eps} \nnb \\
&&\times \Big[\frac{1}{9 \e^5}-\frac{1}{3 \e^4}+\Big(1-\frac{2 \pi^2}{9}\Big) \frac{1}{\e^3}+\Big(-3+\frac{2
\pi^2}{3}-\frac{52}{9} \zeta_3\Big)\frac{1}{\e^2}\nnb\\
&& \hs{15} +\Big(9-2\pi^2+\frac{52}{3}\zeta_3-\frac{7\pi^4}{90}\Big)\frac{1}{\e}\nnb\\
&& \hs{15} +\Big(-27+6\pi^2-52\zeta_3+\frac{7\pi^4}{30}+\frac{68}{9}\pi^2\zeta_3-\frac{140}{3}\zeta_5\Big)\nnb\\
&& \hs{15}
+\Big(81-18\pi^2+156\zeta_3-\frac{7\pi^4}{10}-\frac{68}{3}\pi^2\zeta_3+
140\zeta_5+\frac{2473\pi^6}{34020}+\frac{1136}{9}\zeta_3^2\Big)\e \nnb\\
&& \hs{15}+{\cal{O}}(\e^2)\Big] \; .
\eea

\end{document}